# A Case Study of the Arbitrariness of the h-Index and the Highly-Cited-Publications Indicator

Michael Schreiber

*Institute of Physics, Chemnitz University of Technology, 09107 Chemnitz, Germany. E-mail: schreiber@physik.tu-chemnitz.de*

The arbitrariness of the h-index becomes evident, when one requires $q*h$ instead of $h$ citations as the threshold for the definition of the index, thus changing the size of the core of the most influential publications of a dataset. I analyze the citation records of 26 physicists in order to determine how much the prefactor $q$ influences the ranking. Likewise, the arbitrariness of the highly-cited-publications indicator is due to the threshold value, given either as an absolute number of citations or as a percentage of highly cited papers. The analysis of the 26 citation records shows that the changes in the rankings in dependence on these thresholds are rather large and comparable with the respective changes for the h-index.

**Introduction**

The h-index has been proposed by Hirsch (2005) as a reasonable way to quantify the performance of a researcher by a single number. It is defined as the largest number $h$ of publications which have received at least $h$ citations each. Thus it combines the quantity dimension in terms of the number of publications with the quality dimension in terms of the number of citations. It does not depend explicitly on any parameter. However, as noted by van Eck & Waltman (2008) its definition involves arbitrariness, because one might as well define a generalized index $h_\alpha$ as the highest number of papers that have received at least $\tan(\alpha) * h_\alpha$ citations each. For $\alpha = 45° = \pi/4$ the original definition of the h-index is recovered.

The arbitrariness had already been noted earlier by Lehmann, Jackson, & Lautrup (2006, 2008) as well as Ellison (2010). Ravallion and Wagstaff (2011) regret the missing clear theoretical foundation of the h-index and developed a general approach defining a theoretically consistent measure of influence based on the complete citation curves. For a comparison of the citation curves between different fields an appropriate normalization should be applied (Radicchi, Fortunato, and Castellano, 2008).Waltman & van Eck (2012) argue in favor of the highly-cited-publications indicator rather than the h-index. This has prompted me to study also the arbitrariness of this indicator which I label $H$. For this indicator its arbitrariness is clear, because it depends explicitly on a threshold $p$ that defines which papers are considered to be highly cited or not. The threshold $p$ is given below as a percentage of the total number of papers $N$ together with the corresponding threshold in terms of the number of citations $c_p$ which is needed for a paper in the data base to fall into the respective percentage. Due to the small numbers and the discreteness of the citation distribution, it is not practical in the present investigation to fix a threshold percentage *p* and then determine the threshold number $c_p$ of citations, but rather one should start from a threshold number of citations $c_p$ and then determine the corresponding percentage *p*. This will be done in the analysis below.

Different percentages have been applied, e.g. by Tijssen, Visser, and van Leeuwen (2002), Aksnes (2003), and Albarran, Ortuno, and Ruiz-Castillo (2011). The R6 indicator proposed by Leydesdorff, Bornmann, Mutz, and Opthof (2011) combines 5 different threshold percentages, thus taking the strongly varying skewnesses of various citation distributions into account in one indicator. Similarly Vinkler (2011) couples 14 threshold numbers to construct his CDS index.

The paper is organized as follows: In the next section the definition of the generalized h-index is given and visualized; changes in the ranking for an example of 26 datasets reflecting the citation distributions of 26 physicists are shown and discussed. In the following section the same investigation is performed for the highly-cited-publications indicator. In a final section further discussions and concluding remarks are given.

Admittedly with 26 datasets the present case study is not comprehensive. But it is not the aim of the present investigation to present a statistically reliable analysis for a very large data base. Rather it is the purpose to demonstrate with an example that the arbitrariness of the h-index as well as the highly-cited-publications indicator can strongly mix up the ranking of the datasets, if different thresholds are utilized. Due to the large variety of citation distributions in particular with respect to the different skewness and different tails (for the highly cited papers as well as for the lowly cited papers) it is highly likely that similar mix-ups will usually occur in comparable investigations.

**The Arbitrariness of the h-Index**

First I present the results for the h-index and the generalized index $h_\alpha$. In graphical terms, the determination of the h-index can be visualized by depicting the citation distribution as in Figure 1, where the number of citations $c(r)$ is plotted versus the rank *r* which each paper gets by sorting according to the number of citations. The intersection of these histograms with the diagonal $c(r) = r$ yields the h-indices, i.e. $c(h) = h$. Due to the discreteness of the citation distribution, this condition is not always fulfilled and in order to be precise, one has to determine the largest value of *h* which satisfies

$$h \leq c(h). \qquad (1)$$

This condition implies that $c(h+1) < h+1$.

The slope of the diagonal is given by $q = \tan(\alpha)$ where $\alpha = 45°$ is the angle between the diagonal and the horizontal axis, so that $q = 1$. Thus choosing a different angle $\alpha$ and in that way a different slope $q = \tan(\alpha)$ just means an arbitrary prefactor in the definition of the generalized $h_\alpha$ index, $c(h_\alpha) = \tan(\alpha) h_\alpha = q h_\alpha$ as discussed by Waltman and van Eck (2012). Again, to be precise one has to search for the largest value of $h_\alpha$ which satisfies

$$q h_\alpha \leq c(h_\alpha). \qquad (2)$$

This implies $c(h_\alpha+1) < q(h_\alpha+1)$.

The w-index (Wu, 2010) is an example for α = 84.3°, i.e., a factor $q$ = tan(84.3°) = 10. This is certainly an extreme factor. In a case study of 26 physicists (Schreiber, 2010) I have previously argued that such an extreme factor leads to a large number of tied values of the index, which is not desirable for a ranking. Nevertheless, that analysis showed that even such a large factor does not have a strong influence on the ranking besides the mentioned ties.

It is the purpose of the present investigation to present a case study of the influence of the factor $q$ or, equivalently, the angle α on the ranking. This means that I analyze how changing the size of the core of the most influential publications within all datasets alters the ranking. For that aim I utilize the previously determined citation data of 26 physicists from my home institution (Schreiber, 2008, 2009) which had been determined from the Web of Science in January and February 2007. The datasets are labeled from A to Z from highest to lowest values of $h_{45°}$ as in the previous investigations. The present data base comprises a total number $N$=2373 of publications with a total number $C$=25554 of citations which corresponds to a mean citation rate of 10.77. About 20 % = 473 papers with 15 and more citations have received 17948 citations, i.e., 70 % of all citations, what reflects the skewness of the distribution. The citation distributions of datasets H, J, M and O, P, Q have been presented in Figure 1. It is obvious that choosing another angle instead of the diagonal leads not only to significantly different values of $h_\alpha$ but also to a different ranking of the scientists.

The dependence of the indices $h_\alpha$ on $q$ is shown in Figure 2. Here the datasets have been sorted using the original h-index so that the middle line for α=45°, i.e. $q$=1 is monotonously increasing in the plot. The angle α has been increased and decreased in steps of 5°. Corresponding values for $q$ are denoted in Table 1. This table includes also the accumulated number $n(h_\alpha)$ of papers that belong to all the h-cores[1] for a given α, i.e., how many papers contribute to the 26 $h_\alpha$-indices; this is equivalent to summing the $h_\alpha$ values for each α.

One can see in Figure 2 that already for α = 50° there are small fluctuations which destroy the monotony. For larger and larger values of α and $q$ the fluctuations increase and appear for different datasets. This means that the ranking is changed frequently. These changes reflect the different skewnesses of the 26 citation distributions, giving more weight to more strongly skewed distributions. Whether this influence is desirable remains questionable.

The respective effects are much stronger for smaller values of $q$ and α decreasing towards 0. Here the lowly cited publications in the long tail of the citation distributions become more and more important. This is strongly influenced by different publication strategies and I do not think that relatively low values of $q$ and α are reasonable, because in my opinion they overvalue the lowly cited papers.

The changes in the ranking are visualized in Figure 3. Although the fluctuations in Figure 1 appear rather small at least for large values of α, the ranks fluctuate strongly for small α as well as for large α. Only at the top of the ranking nothing changes, and at the bottom the changes are small and often restricted to tied ranks instead of interchanged ranks. A noteworthy exception is the dataset Z which remains at rank 26 or 25.5 for α = 5° up to α = 70° and then advances to rank 20.5, because it is so strongly skewed. In the center of the ranking the changes are sometimes dramatic. For example datasets H and I advance along different paths from rank 15.5 for $h_{5°}$ to rank 6.5 for $h_{85°}$. Researcher J rises from 21 for $h_{5°}$ to 8.5 before dropping again to 14.5 and ending at rank 13.5 for $h_{85°}$. On the other hand scientists M and N start at rank 7 for $h_{5°}$ and drop along different paths to rank 13.5 for $h_{85°}$. The strongest drop occurs for scientist O from 10.5 to 20.5, the largest improvement for researcher P from 22.5 to 6.5 and it is also interesting to note that these big changes are far from monotonous.

---

[1] For each dataset the h-core comprises those papers which contribute to the h-index.

In order to quantify these changes, Pearson's correlation coefficients κ and Spearman's rank-order correlation coefficients $κ_s$ between the original h-index and the $h_α$ values are presented in Table 1. In principle, when comparing rank orders, one should use Spearman's correlation coefficients. However, as there are several studies which have utilized Pearson analysis, I also show the respective values in Table 1. The different results are due to the fact that rather small differences in the index values can lead to large deviations of the ranks especially if several datasets are tied, i.e. have identical index values and thus the same rank. The resulting gaps in the rank order, which can be clearly seen on the right edge of Figure 3 where 21 datasets in the center of the figure are tied at three rank values, reduce Spearman's correlation coefficients strongly. But it appears doubtful to me whether this is meaningful, because the small changes in the index values should not be overinterpreted.

For the Pearson analysis it turns out that the correlations are rather strong, with correlation coefficients close to 1 and smaller than 0.97 only for the extreme cases of α < 15° or α > 75°. Even the rank-order correlation coefficients are close to 1 in most cases and remain larger than 0.95 except for α ≤ 15° or α ≥ 75°.

In conclusion, exploiting the arbitrariness of the h-index leads to drastic changes in the ranking, but surprisingly this is not reflected strongly in the correlation coefficients. In order to test whether this unexpected result might be caused by the relatively stable ranks of the top and bottom datasets, I have repeated the analysis for datasets D – W. The calculated correlation coefficients for this subset are much smaller, see Table 1. This means that the rather large correlation coefficients for the complete set are somewhat misleading.

**The Arbitrariness of the Highly-Cited-Publications Indicator**

Now I turn to the highly-cited-publications indicator *H*. The indicator *H* (written in italics) should not be confused with the scientist H (written not in italics). As a starting point I define the number of highly cited papers as the top *p* = 16.4 %. The objective behind this choice is that the respective threshold $c_p$ = 18 citations corresponds to a total number $n_p$ = 389 papers which is as close as possible to the accumulated number of papers $n(h_{45°})$ = 387 in all the 26 h-cores for the original h-index, see Table 1. The resulting values of the indicator *H* are compared with the h-indices in Table 2. One observes that the absolute values of *H* are spread over a larger range than those for *h*. It is obvious that a different ranking as compared to the h-index values occurs. The ranking in the middle of the sample is somewhat changed, with the largest drop of 4 ranks for scientist Q and the largest advance by 4 ranks for researcher U but also a conspicuous advance of 3 steps for the scientist Z, especially as this dataset achieved the smallest h-index and thus the lowest rank in the h-index list. The relatively small differences can be quantified, Pearson's correlation coefficient κ between the h-index values and the $H$ values is 0.965, and Spearman's rank-order correlation coefficient $κ_s$ is 0.979.

In order to study the arbitrariness of the indicator $H$ I define a generalized indicator $H_p$ for which I have changed the threshold; the chosen values for $c_p$ and the corresponding values of $p$ are given in Table 3. Here again the objective in each case has been to get a resulting accumulated number $n(H_p)$ of highly-cited-publications for all the 26 datasets which is as close as possible to the accumulated sizes $n(h_α)$ of the h-cores as given in Table 1. The resulting values of $n(H_p)$ are presented in Table 3.

In Figure 1 the horizontal lines indicate those 5 thresholds $c_p$, for which $n(H_p)$ is close to the respective values of $n(h_α)$ for the dashed lines. Again it is obvious from Figure 1 that changing the threshold $c_p$ leads not only to different values of $H_p$, but also to different rankings.

It is interesting to note that the intersections of the horizontal lines and the dashed lines are always above the histograms for the 6 datasets presented in Figure 1. This means that the $H_p$ values are always smaller than the corresponding $h_α$ values for these 6 scientists. A closer inspection of the raw data showed that this is usually the case except for the top scientists. For the top three scientists the $H_p$ values are much larger than the $h_α$ values so that the overall equivalence of $n(H_p)$ and $n(h_α)$ is achieved.

The thresholds 15 % and 10 % have been utilized previously. For the complete data base this means 356 and 237 papers, respectively. I note that $c_p = 19$ corresponds to $p = 14.8 ≈ 15$ and $c_p = 26$ corresponds to $p = 9.9 ≈ 10$, which in turn means that 351 and 236 papers are taken into consideration for these cases.

The effect of varying the threshold according to Table 3 is visualized in Figure 4. Here the datasets have been ordered by the indicator $H$, so that the middle line for $c_p = 18$ citations, i.e., for $p = 16.4$ % is monotonously increasing in the plot. The overall impression is very similar to Figure 2. Small fluctuations occur destroying the monotony. For extreme values of $p$ the fluctuations are stronger.

Figure 5 shows the changes in the ranking in dependence on the threshold. Increasing the threshold from 5 to 51 citations, researcher P drops from rank 19.5 to 21.5 before advancing to rank 5, and I rises from 13.5 to rank 5, too. Also, X improves from rank 24 to 15.5. More conspicuous is the case H advancing from 15 to 8, before dropping to 10 and rising again to 5.5 but dropping once more to end up at rank 7.5. On the other hand, the largest drops occur for scientist O from position 8 to 22.5 and Q from 12 to 22.5, too. Strong fluctuations can again be seen for N, dropping from rank 6 to 14, but advancing to 8.5 before dropping again to 15.5. Very strong fluctuations can be observed for researcher G but also for M, although for both cases their initial and final ranks are not very different.

Thus, like Figure 4 appears similar to Figure 2, for Figure 5 the overall impression is also similar to Figure 3 and it is difficult to decide in which figure the disarray is stronger. The answer is not even clearly given by the correlation coefficients κ and the rank-order correlation coefficients $κ_s$ in Table 3: One might conclude that the disarray is comparable, because most of the values for $H_p$ are rather similar to those for $h_\alpha$ (with a notable exception of the $\kappa_s$ for large α and large $c_p$), and no clear trend is observable. Accordingly like for $h_\alpha$, the correlation coefficients and the rank-order correlation coefficients for $H_p$ are surprisingly large in view of the strong fluctuations in Figure 5. And once more a restriction to 20 datasets by excluding the top 3 and bottom 3 datasets leads to significantly smaller correlation coefficients, see also Table 3. This means that like for the generalized h-index the relatively stable ranks of the top and bottom datasets have a strong influence on the correlation coefficients, which could lead to the wrong impression that the changes are not so substantial.

**Discussion and Conclusions**

In conclusion, the present case study has shown how the arbitrariness of the h-index can influence the ranking, in particular, when rather extreme values are chosen for the proportionality factor $q$, or, equivalently for the angle α of the line intersecting the histogram of the citation distributions in the graphical representation. This means that the choice of $q$ or α will randomly favor or disfavor individuals, as visualized in Figures 2 and 3.

The highly-cited-publications indicator needs a threshold number of citations for its definition, so that its arbitrariness is obvious. Therefore, Hirsch (2005) argued against this indicator because that threshold "will randomly favor or disfavor individuals" (p. 16569). Figures 4 and 5 demonstrate that there is some truth in this criticism. The fluctuations are similar in comparison to the generalized h-index, compare Figures 2 and 4. This is true also for the ranks, compare Figures 3 and 5.

In conclusion, it is important how the threshold is determined. The Essential Science Indicators from Thomson Reuters select the top 1% of articles as highly cited papers. In the present case this would mean $n(H_p) = 24$ papers, corresponding to a threshold of $c_p = 104$ citations. Such a small number of highly cited papers is also not practical in the present case, where 26 datasets are to be compared. Tijssen, Visser, and van Leeuwen (2002) have analyzed the top 1% and the top 10% highly cited papers in the Netherlands from 1980 - 1989. With regard to the $H_p$ index they observed that "the statistical robustness of this index diminishes as the number of papers decline ... especially below the level of 100" (p. 387). In that study certain time intervals were analyzed, so that relatively few highly cited papers remain in the sample in a given interval. In the present investigation this is irrelevant, but it emphasizes that it is dangerous to define the threshold too high, because then too few highly cited papers are taken into account. Aksnes (2003) used 17 times the mean citation rate as a threshold for the distinction of highly cited papers in a study of Norwegian scientists. In the present case this would mean $c_p = 17*10.77 = 183.1$ which would only leave $n(H_p) = 5$ highly cited papers and would therefore be not very useful. For the present dataset a value of twice the mean citation rate, i.e. $c_p = 21.5$ seems to be a reasonable choice, yielding $n(H_p) = 294$ highly cited papers, i.e. 12.4 %.

Ellison (2010) has discussed variants of the Hirsch index and concluded that those variants that emphasize smaller numbers of highly cited papers perform better than the original index. For $h_\alpha$ in a comparison with labor market outcomes of economists, he concluded that $q$ should be substantially larger than one and found the best result for $q = 10$. In the present investigation this would mean $\alpha = 84.3°$ and it would yield $n(h_\alpha) = 92$ publications. This relatively small number results in so many ties that only the top 3 scientists are clearly distinguished, while 2, 5, 6, 10 scientists have tied ranks with $h_{84.3°} = 1, 2, 3, 4$, respectively. Thus this extreme value of $q$ is not very useful for a ranking in the present case. Rather the original index $h = h_{45°}$ seems to be a reasonable choice, although it also leaves 14 scientists in 5 ties as shown in Table 2.

Whether this is also the best choice remains an open question. Following the above argumentation, I conclude that large values of $q$ should not be used. Small values of $q$ put more emphasis on the long tail of lowly cited publications and are therefore also not suitable. I think that $q$ values between 1/3 and 3 are reasonable and it depends on the evaluator, whether larger values are favored or not. In particular for top scientists with a much more skewed citation record larger values of $q$ might be more practical. In any case, it is important to accept that contrary to common belief there is an arbitrariness involved in the definition of the h-index. The prefactor $q = 1$ in the original definition has become a standard. But this does not mean that it is better than say $q = ½$ or $q = 2$.

One cannot define a best value of $q$ for all evaluations. But I propose to utilize a value of $q = 3$ for comparing eminent scientists, because such a relatively large $q$ value would reasonably reduce the size of their h-cores which is unnecessarily large for the original definition with $q = 1$. The practical advantage is that the precision problem is reduced, because less papers have to be checked. For a group of more average scientists like in the present case study, the standard $q = 1$ is reasonable. I think that larger values of $q$ would reduce too much the number of papers which are considered to be relevant. For an evaluation of a group of junior scientists I propose to use a value of 1/3 in order to achieve a better distinguishability by increasing the size of the h-cores. In my opinion, even smaller values of $q$ would enhance the danger that irrelevant differences are overinterpreted.

Likewise, for the highly-cited-publications indicator it is not possible to determine a best threshold. Again it depends on the evaluator whether the top cited papers should be more or less favored. The 10% threshold has become one standard and as a relative threshold this appears to be a reasonable choice in the present investigation, too, corresponding to a citation number threshold of 26. However, usually one should compare the citation distributions in this approach with a large reference set. Using all physics papers published between 1986 and 2010, the citation number threshold of the top 10% cited publications turns out to be 37 (Schreiber, 2012). In the present case this would mean only 6.36%, namely 151 top cited papers. As discussed above, such a small number leads to many ties and such a reduced distinguishability is not attractive. In contrast to the present investigation, Albarran, Ortuno, and Ruiz-Castillo (2011) concluded that changing the threshold percentage from 20% to 5% has relatively small impact on the relative positions of the three investigated geographical areas.

In the present investigation I have determined the thresholds not with respect to a large reference set as is the common standard for percentile-based indicators. I have rather used only a unified set of the 26 example datasets as the reference set for determining the absolute threshold number of citations in Tables 2 and 3 and the corresponding threshold percentages defining the core of most influential publications. In this way one can avoid the difficulty to obtain the characteristics of the citation distributions for a large reference set. This simplifies the calculation of $H_p$ considerably. Likewise, for the determination of $h_\alpha$ no reference set (not even a unified set) is necessary because each subset defines its own citation threshold, and this feature has certainly contributed to the attractiveness of the h-index. Another method to calculate thresholds without referring to reference sets or without even referring to a unified set would be to determine the core of most influential publications implicitly for each subset, e.g. the application of an elite set of papers defined by the square root of the number of publications in a subset (Vinkler, 2009).

The erratic fluctuations of $H_p$ in Figure 4 and of the corresponding rather chaotic changes of the ranks in Figure 5 might be tolerated, if the threshold is indeed determined by the absolute number of citations. But if a percentage is used as threshold, an inconsistency problem arises, because that percentage threshold can be influenced by the inclusion or deletion of a single researcher from the dataset. For example, excluding scientist A from the current analysis reduces the total number of publications by 12.2 % to $N = 2083$. Excluding A and B leaves $N = 1813$ papers with $C = 16380$ citations. Of course, now several highly cited papers are excluded from the data base. Thus the percentage of highly cited papers for a given citation number threshold changes significantly, as shown above in the discussion of Table 3. The other way round, to achieve a percentage threshold of 10 %, one needs now a lower citation number threshold of 23 instead of 26. Likewise, a percentage threshold of 15 % now means to take into account papers with only at least 17 instead of 19 citations. Consequently the ranking of the remaining 24 scientists changes, in some cases dramatically. Note that the data for the latter case ($p = 15$, $c_p = 19$ and $c_p = 17$) are contained in Figures 4 and 5 so that a visual comparison is possible. In fact, one observes that 19 of the now 24 datasets change ranks; G and O drop most strongly by 2 ranks, and I advances 2.5 steps, only because A and B have been excluded from the reference set.

Such an inconsistency does not occur for the generalized h-index, because of its implicit definition. However, as noted by Waltman & van Eck (2012) other inconsistencies can be observed for the h-index and it thus remains an individual choice whether the h-index or the highly-cited-publications indicator $H$ is considered to be more attractive. Many studies have criticized the h-index and a large number of variants have been invented, e.g. normalization with respect to different disciplines (Radicchi, Fortunato, and Castellano, 2008). The arbitrariness of $h$ which has been discussed in the present investigation applies to most, if not all of these variants and does not help solving difficulties with the statistics of the h-index and other of its shortcomings, nor does it help selecting one of the variants in favor of another.

**References**

**Aksnes, D.W. (2003).** Characteristics of highly cited papers. Research Evaluation 12(3), 159-170.

**Albarrán, P., Ortuño, I., & Ruiz-Castillo, J. (2011).** High- and low-impact citation measures: Empirical applications. Journal of Informetrics 5, 122-145.

**Ellison, G. (2010).** How does the market use citation data? The Hirsch index in economics. (CESifo Working Paper 3188) Massachusetts Institute of Technology, Cambridge, USA. http://www.cesifo-group.de/portal/pls/portal/docs/1/1185252.PDF

**Hirsch, J.E. (2005).** An index to quantify an individual's scientific research output. Proceedings of the National Academy of Science, 102(46), 16569-16572.

**Lehmann, S., Jackson, A.D., & Lautrup, B.E. (2006).** Measures for measures. Nature, 444, 1003-1004.

**Lehmann, S., Jackson, A.D., & Lautrup, B.E. (2008).** A quantitative analysis of indicators of scientific performance. Scientometrics, 76(2), 369-390.

**Leydesdorff, L., Bornmann, L., Mutz, R., & Opthof, T. (2011).** Turning the tables on citation analysis one more time: Principles for comparing sets of documents. Journal of the American Society for Information Science and Technology, 62(7), 1370–1381.

**Radicchi, F., Fortunato, S., & Castellano, C. (2008).** Universality of citation distributions: Toward an objective measure of scientific impact. Proceedings of the National Academy of Science 105(45), 17268-17272.

**Ravallion, M. & Wagstaff, A. (2011).** On measuring scholarly influence by citations. Scientometrics, 88, 321-337.

**Schreiber, M. (2008).** An empirical investigation of the *g*-index for 26 physicists in comparison with the *h*-index, the *A*-index, and the *R*-index. Journal of the American Society for Information Science and Technology, 59(9), 1513-1522.

**Schreiber, M. (2009).** A case study of the modified Hirsch index $h_m$ accounting for multiple coauthors. Journal of the American Society for Information Science and Technology, 60(6), 1274-1282.

**Schreiber, M. (2010).** Twenty Hirsch index variants and other indicators giving more or less preference to highly cited papers. Annalen der Physik, 552(8), 536-554.

**Schreiber, M. (2012).** How much do different ways of calculating percentiles influence the derived performance indicators? – A case study. Scientometrics (accepted).

**Tijssen, R.J.W., Visser, M.S., & van Leeuwen, T.N. (2002).** Benchmarking international scientific excellence: Are highly cited research papers an appropriate frame of reference?. Scientometrics, 54(3), 381-397.

**Van Eck N.J. & Waltman, L. (2008).** Generalizing the h- and g-indices. Journal of Informetrics, 2(4), 263-271.

**Vinkler, P. (2009).** The π-index: a new indicator for assessing scientific impact. Journal of Information Science, 35(5), 602-612.

**Vinkler, P. (2011).** Application of the distribution of citations among publications in scientometric evaluations. Journal of the American Society for Information Science and Technology, 62(10), 1963-1978.

**Waltman, L. & van Eck N.J. (2012).** The inconsistency of the h-index. Journal of the American Society for Information Science and Technology, 63(2), 406-415.

**Wu, Q. (2010).** The w-index: A measure to assess scientific impact by focusing on widely cited papers. Journal of the American Society for Information Science and Technology, 61(3), 609-614.

TABLE 1. Pearson's correlation coefficients κ and Spearman's rank-order correlation coefficients $\kappa_s$ between the original h-index and the $h_\alpha$ values for the data shown in Figures 2 and 3; the correlation coefficients after excluding the top 3 and the bottom 3 datasets from the sample are also given; all correlation coefficients are significant at the 0.001 level as determined by the t distribution, except the value marked with * which is significant at the 0.01 level. $q$ is the slope of the $h_\alpha$-determining line (compare Fig. 1) and $n(h_\alpha)$ is the accumulated number of papers in all h-cores for each value of α.

| α | 5° | 10° | 15° | 20° | 25° | 30° | 35° | 40° | 45° | 50° | 55° | 60° | 65° | 70° | 75° | 80° | 85° |
|---|---|---|---|---|---|---|---|---|---|---|---|---|---|---|---|---|---|
| $q=\tan\alpha$ | 0.09 | 0.18 | 0.27 | 0.36 | 0.47 | 0.58 | 0.70 | 0.84 | 1.00 | 1.19 | 1.43 | 1.73 | 2.14 | 2.75 | 3.73 | 5.67 | 11.43 |
| $n(h_\alpha)$ | 1119 | 875 | 730 | 636 | 566 | 505 | 454 | 413 | 387 | 339 | 305 | 271 | 240 | 210 | 174 | 133 | 83 |
| κ(A-Z) | 0.923 | 0.946 | 0.971 | 0.977 | 0.984 | 0.991 | 0.996 | 0.998 | 1.000 | 0.993 | 0.993 | 0.989 | 0.991 | 0.987 | 0.972 | 0.968 | 0.949 |
| $\kappa_s$(A-Z) | 0.829 | 0.888 | 0.928 | 0.958 | 0.967 | 0.975 | 0.986 | 0.992 | 1.000 | 0.984 | 0.993 | 0.980 | 0.981 | 0.977 | 0.928 | 0.947 | 0.893 |
| $\kappa_s$(D-W) | 0.619* | 0.753 | 0.842 | 0.907 | 0.927 | 0.944 | 0.968 | 0.982 | 1.000 | 0.965 | 0.984 | 0.957 | 0.957 | 0.950 | 0.844 | 0.902 | 0.790 |

TABLE 2. Values of the h-index and the highly cited publications indicator H for the 26 datasets, where a threshold $p = 16.4\%$ corresponding to $c_p = 18$ is utilized; the datasets are sorted using the h-index; the ranks determined from the h-index and the highly cited publications indicator are also given. For tied papers the average rank is given.

| dataset | A | B | C | D | E | F | G | H | I | J | K | L | M |
|---|---|---|---|---|---|---|---|---|---|---|---|---|---|
| h-index | 39 | 27 | 23 | 20 | 19 | 18 | 17 | 16 | 15 | 15 | 14 | 14 | 14 |
| rank(h) | 1 | 2 | 3 | 4 | 5 | 6 | 7 | 8 | 9.5 | 9.5 | 12.5 | 12.5 | 12.5 |
| indicator H | 88 | 49 | 32 | 28 | 20 | 18 | 15 | 13 | 14 | 13 | 12 | 13 | 10 |
| rank(H) | 1 | 2 | 3 | 4 | 5 | 6 | 7 | 10 | 8 | 10 | 12 | 10 | 13 |

| Dataset | N | O | P | Q | R | S | T | U | V | W | X | Y | Z |
|---|---|---|---|---|---|---|---|---|---|---|---|---|---|
| h-index | 14 | 13 | 13 | 13 | 12 | 12 | 10 | 10 | 10 | 9 | 8 | 7 | 5 |
| rank(h) | 12.5 | 16 | 16 | 16 | 18.5 | 18.5 | 21 | 21 | 21 | 23 | 24 | 25 | 26 |
| indicator H | 9 | 8 | 8 | 5 | 6 | 6 | 4 | 7 | 4 | 2 | 1 | 1 | 3 |
| rank(H) | 14 | 15.5 | 15.5 | 20 | 18.5 | 18.5 | 21.5 | 17 | 21.5 | 24 | 25.5 | 25.5 | 23 |

TABLE 3. Pearson's correlation coefficients κ and Spearman's rank-order correlation coefficients $\kappa_s$ between the original highly cited publications indicator H and the $H_p$ values for the data shown in Figures 4 and 5; the correlation coefficients after excluding the top 3 and the bottom 3 datasets from the sample are also given; all correlation coefficients are significant at the 0.001 level as determined by the t distribution, except the value marked with * which is significant at the 0.01 level. $p$ is the percentage of highly cited papers corresponding to the threshold number $c_p$ of citations and $n(H_p)$ is the accumulated number of highly cited papers in all 26 datasets for each value of $c_p$.

| $c_p$ | 5 | 8 | 9 | 11 | 13 | 14 | 16 | 17 | 18 | 19 | 21 | 24 | 26 | 29 | 32 | 40 | 51 |
|---|---|---|---|---|---|---|---|---|---|---|---|---|---|---|---|---|---|
| $p$ | 49.6 | 36.6 | 33.0 | 27.6 | 23.6 | 21.9 | 18.8 | 17.6 | 16.4 | 14.8 | 12.9 | 11.1 | 9.95 | 8.60 | 7.29 | 5.60 | 3.50 |
| $n(H_p)$ | 1178 | 868 | 782 | 654 | 559 | 520 | 447 | 417 | 389 | 351 | 307 | 264 | 236 | 204 | 173 | 133 | 83 |
| κ(A-Z) | 0.932 | 0.955 | 0.962 | 0.979 | 0.985 | 0.990 | 0.997 | 0.998 | 1.000 | 0.997 | 0.993 | 0.990 | 0.989 | 0.983 | 0.975 | 0.977 | 0.968 |
| $\kappa_s$(A-Z) | 0.846 | 0.936 | 0.931 | 0.955 | 0.974 | 0.983 | 0.996 | 0.994 | 1.000 | 0.994 | 0.977 | 0.963 | 0.944 | 0.944 | 0.891 | 0.891 | 0.836 |
| $\kappa_s$(D-V,Z) | 0.666* | 0.863 | 0.853 | 0.908 | 0.945 | 0.966 | 0.993 | 0.986 | 1.000 | 0.987 | 0.953 | 0.924 | 0.904 | 0.905 | 0.791 | 0.823 | 0.760 |

FIG 1. Citation distributions of 6 scientists, i.e., number of citations versus paper number after sorting the papers according to the number of citations; the 5 dashed lines are plotted at angles of 75°, 60°, 45°, 30°, 15° with the horizontal axis. The horizontal lines reflect thresholds of $c_p$ = 9, 14, 18, 24, 32 citations for $H_p$.

FIG 2. Dependence of the generalized indices $h_\alpha$ on $q = \tan(\alpha)$ for the 26 datasets in the present investigation. $q$ increases from $q = 0.09$ (top) to 11.43 (bottom).

FIG 3. Ranking of the 26 scientists in dependence on α, determined from the $h_\alpha$ values in Figure 2; for tied ranks the average is given; various segments for some curves are slightly shifted by ±0.1 or ±0.2 in order to facilitate distinguishing piled-up segments; thicker lines are used for datasets which are mentioned in the text. These datasets are marked with filled symbols, the other datasets are marked with line symbols.

FIG 4. Same as Figure 2, but for the highly-cited-publications indicator $H_p$ in dependence on the threshold $p$. Note the different sequence of the datasets in comparison to Figure 2 as indicated on the horizontal axis. $p$ increases from $p = 5$ (top) to 51 (bottom).

FIG 5. Same as Figure 3, but for the values of the highly-cited-publications indicator $H_p$ presented in Figure 4; for tied ranks the average is given; various segments for some curves are slightly shifted by ±0.1 or ±0.2 in order to facilitate distinguishing piled up segments; thicker lines are used for datasets which are mentioned in the text. These datasets are marked with filled symbols, the other datasets are marked with line symbols.

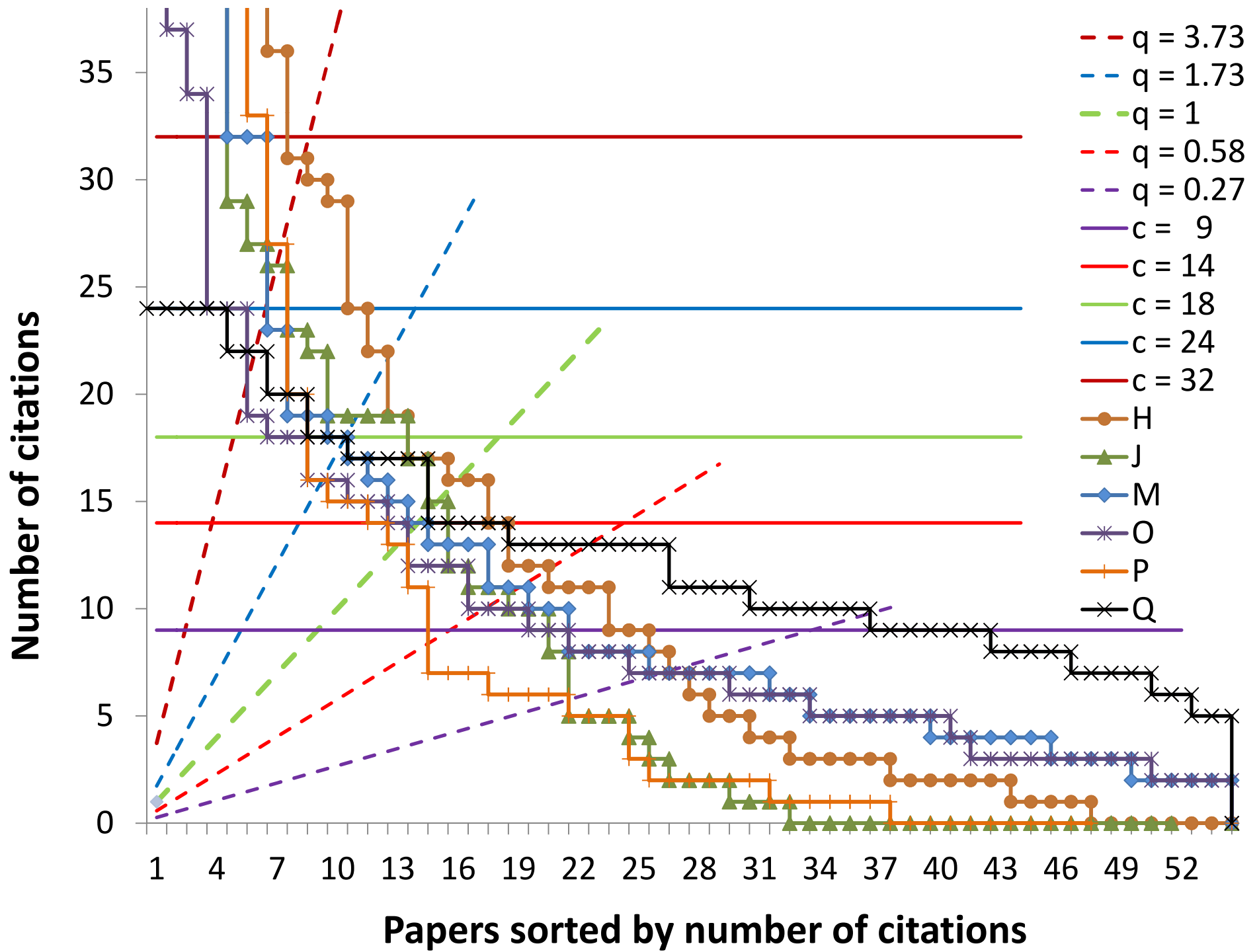

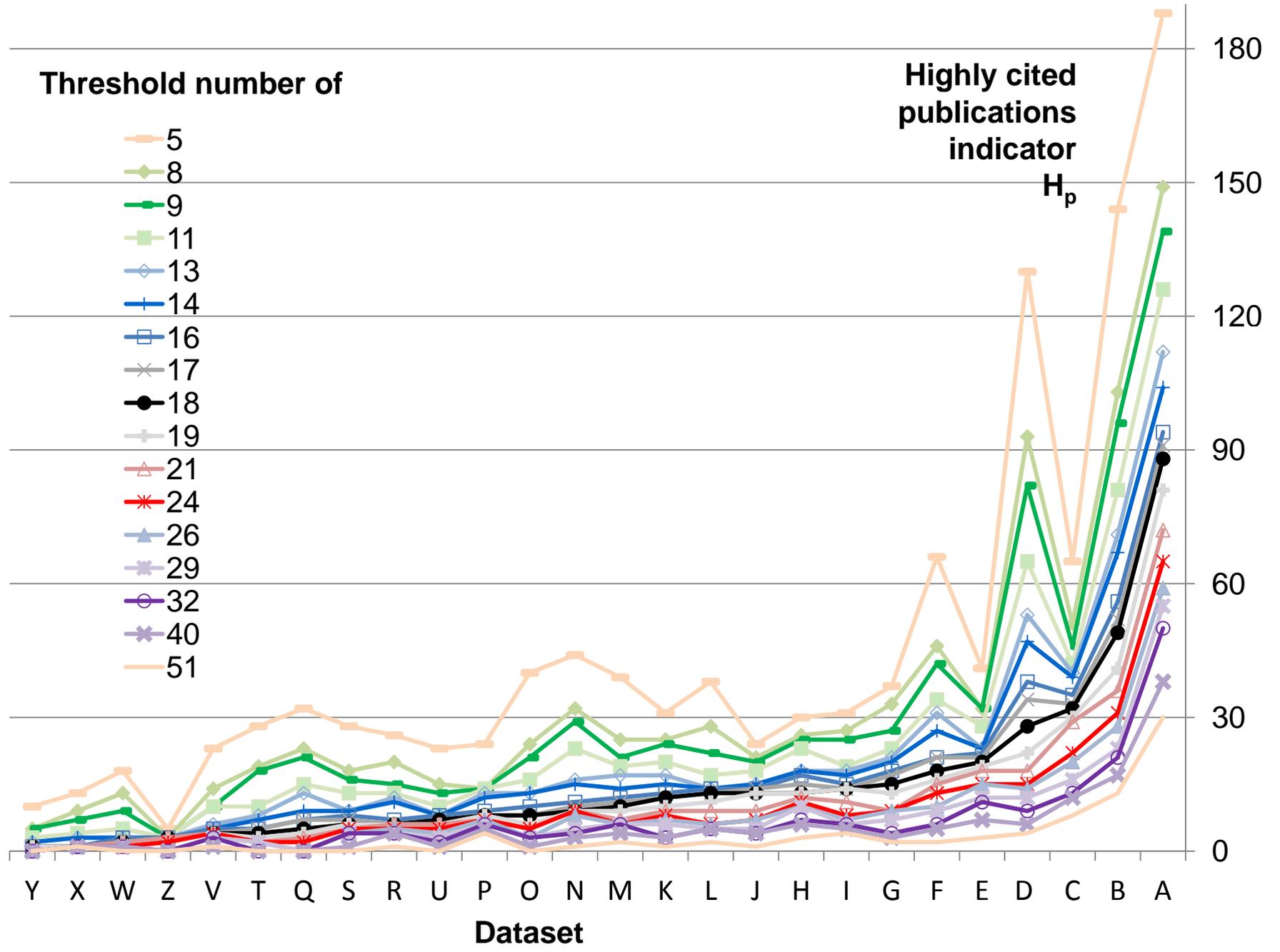